\documentstyle[epsf]{article}

\begin{document}
\mbox{ }
\rightline{UCT-TP-259/01}\\
\rightline{February 2001}\\
\vspace{3.5cm}
\begin{center}
{\Large \bf Pion form factor in large $N_{c}$  QCD}\footnote{Work supported
in part by the Volkswagen Foundation}\\
\vspace{.5cm}
{\bf C. A. Dominguez}\\[.5cm]
Institute of Theoretical Physics and Astrophysics\\
University of Cape Town, Rondebosch 7700, South Africa\\[.5cm]
\end{center}
\vspace{.5cm}
\begin{abstract}
\noindent
The electromagnetic form factor of the pion is obtained using a particular
realization of QCD in the large $N_c$ limit, which sums up the infinite
number of zero-width resonances to yield an Euler's Beta function of the
Veneziano type. This form factor agrees with space-like data much better than single
rho-meson dominance. A simple unitarization
ansatz is illustrated, and the resulting vector spectral function in the
time-like region is shown to be in  reasonable agreement with the ALEPH data
from threshold up to about 1.3 $\mbox{GeV}^2$.
\end{abstract}
\newpage
\setlength{\baselineskip}{1.5\baselineskip}
\noindent

It is well known that QCD in the limit of large number of colours
($QCD_{\infty}$) \cite{GTH} predicts the existence of an
infinity of zero-width resonances \cite{W}. However, in the absence of
exact solutions to this theory, the hadronic parameters (masses, widths,
couplings) remain unspecified. Various models of this spectrum have been
proposed for heavy quark Green's functions \cite{SH}-\cite{CAD1}, as well
as for light quark systems \cite{DR}. An infinite set of narrow resonances
is reminiscent of Veneziano's dual model \cite{VEN}; in fact, such a
connection has been studied recently for the case of vector and axial-vector
two-point functions \cite{SP}. In this note the three-point function
determining the electromagnetic form factor of the pion is analyzed using a
specific realization of $QCD_{\infty}$, namely a choice of
masses and couplings that yields an Euler's  Beta function of the Veneziano
type. This realization will be called  Dual-$QCD_{\infty}$ in the sequel.
Models of this kind have been studied long before the advent of
$QCD_{\infty}$, in connection with electromagnetic form factors \cite{FRA},
purely hadronic three-point functions \cite{CAD2}-\cite{CAD3}, as well as
two-point functions \cite{CAD4}. However,
the connection to $QCD_{\infty}$ was lacking. The main purpose of this
note is to show that the particular choice of masses and couplings leading
to a Beta function form factor is a very viable way of
modelling $QCD_{\infty}$. In fact, this form factor exhibits asymptotic Regge
behaviour in the space-like region, where it agrees with experiment much
better than single rho-meson dominance. Also, unlike four-point functions,
this form factor can be unitarized without unwanted consequences.
A simple unitarization ansatz will be illustrated. This leads to a
smeared vector spectral function in the time-like region 
which can then be confronted, both locally and globally,
with the ALEPH data \cite{ALEPH}. \\

With the standard definition of the pion form factor
\begin{equation}
< \pi(p_{2}) | J_{\mu}^{EM} | \pi(p_{1})> = (p_{1}+p_{2})_{\mu} \;\; F_{\pi}
(s) \; ,
\end{equation}
where $s \equiv q^{2}=(p_{2}-p_{1})^{2}$, one expects in $QCD_{\infty}$
\begin{equation}
F_{\pi}(s) = \sum_{n=0}^{\infty}
\frac{C_{n}}{(M_{n}^{2} -s)} \; .
\end{equation}
In order to obtain a Beta function type  form factor the
coefficients $C_{n}$ are chosen as
\begin{equation}
C_{n} = \frac{\Gamma(\beta-1/2)}{\alpha' \sqrt{\pi}} \; \frac{(-1)^n}
{\Gamma(n+1)} \;
\frac{1}{\Gamma(\beta-1-n)} \; ,
\end{equation}
where $\beta$ is a free parameter to be fit e.g. from data in the
space-like region ($s<0$), and $\alpha' = 1/2 M_{\rho}{^2}$ is the universal
string tension entering the rho-meson Regge trajectory
\begin{equation}
\alpha_{\rho}(s) = 1 + \alpha ' (s-M_{\rho}^{2}) \; .
\end{equation}
On the other hand, the mass spectrum is chosen as \cite{AB}
\begin{equation}
M_{n}^{2} = M_{\rho}^{2} (1 + 2 n) \; .
\end{equation}
These choices then lead to
\begin{eqnarray}
F_{\pi}(s) &=& \frac{\Gamma(\beta-1/2)}{\sqrt{\pi}} \; \sum_{n=0}^{\infty}\;
\frac{(-1)^{n}}{\Gamma(n+1)} \; \frac{1}{\Gamma(\beta-1-n)} \; \frac{1}
{[n+1-\alpha_\rho(s)]} \nonumber \\ [.2cm]
& = &
\frac{1}{\sqrt{\pi}} \; \frac{\Gamma (\beta-1/2)}{\Gamma(\beta-1)}   \;\;
B(\beta - 1,\; 1/2 - \alpha' s) \;,
\end{eqnarray}
where B(x,y) is Euler's Beta function, and $F_\pi(0) = 1 $.
In the time-like region ($s>0$) the
poles of the Beta function along the negative real axis correspond to an
infinite set of zero-width resonances with equally spaced squared masses
given by Eq.(5). Asymptotically, the pion form factor
in the space-like region exhibits Regge behaviour, viz.
\begin{equation}
\lim_{s \rightarrow - \infty} F_\pi(s) = (- \alpha' \;s)^{(1-\beta)}  \; ,
\end{equation}
which can be used to fix the parameter $\beta$ from a fit to the data.
Alternatively, $\beta$ can be determined from the measured mean squared
radius of the pion. As shown below, both procedures are nicely consistent.
It should be noticed that the value $\beta = 2$ reduces the form factor
to  single rho-meson dominance (naive Vector Meson Dominance).\\

The mass formula Eq.(5) predicts e.g. for the first two radial excitations:
$M_{\rho'} \simeq 1340$ MeV, and $M_{\rho''} \simeq 1720$ MeV,
in reasonable agreement with experiment
\cite{PDG} : $M_{\rho'} = 1465 \pm 25$ MeV, and  $M_{\rho''} = 1700 \pm 20$
MeV. The form factor Eq. (6) is shown in Fig. 1 (solid line),
together with the available experimental data in the space-like
region \cite{DATA}-\cite{AMEN} (including the latest data from the
Jefferson Lab), and for a least-squared fitted value $ \beta \simeq 2.3$.
The Vector Meson Dominance (VMD) prediction ($\beta = 2$) is also shown for
comparison (dash line). The statistical significance of the difference
between both
predictions is illustrated by the resulting chi-squared values. In the
case of $\beta = 2.3$, implying an infinite set of vector mesons,
the chi-squared per degree of freedom is $\chi^2_F \simeq 1.4$, while
VMD ($\beta=2$) gives $\chi^2_F \simeq 11$, or an unacceptably poor fit to
the data. On the other hand, the mean-squared radius which follows from
Eq.(6) is $<r^2_\pi> = 0.436 \;\mbox{fm}^2$, to be compared with the
experimental value
$<r^{2}_{\pi}> = 0.439 \pm 0.008 \; \mbox{fm}^{2}$ \cite{AMEN},
and that of VMD ($\beta = 2$): $<r^2_\pi> = 0.394 \;\mbox{fm}^{2}$.\\

Dual-$QCD_{\infty}$ may also be viewed as a particular realization of
Extended VMD \cite{AB}, according to which one expects in general
\begin {equation}
F_\pi(s) = \sum_{n=0}^{\infty} \; \frac{g_{\rho_n \pi\pi}}{f_n}
\; \frac{M_n^2}{(M_n^2 -s)} \; ,
\end{equation}
where the $f_n$ are the electromagnetic couplings of the photon to the
vector mesons $\rho_n$ of masses $M_n$, and $g_{\rho_n \pi\pi}$ are the
$\rho_n \pi\pi$ strong couplings. In fact, in the framework of
Dual-$QCD_\infty$, Eq.(6) may be rewritten as
\begin {equation}
F_\pi(s) = \frac{g_{\rho \pi \pi}}{f_\rho}\;
\frac{M_\rho^2}{(M_\rho^2 -s)} \; F_{\rho \pi \pi}(s) \; ,
\end{equation}
where
\begin {equation}
F_{\rho \pi \pi}(s) = \frac{\Gamma (\beta-1)}{\Gamma(\beta-2)} \;
B(\beta - 2, \;3/2 - \alpha' s) \; ,
\end {equation}
is the $\rho\pi\pi$ vertex function with an of-mass-shell rho-meson,
normalized as $F_{\rho \pi \pi}(s=M_\rho^2) = 1$.
Once again, for $\beta=2$ one recovers the VMD result. From the normalization
$F_\pi(0) = 1$, there follows the VMD universality relation
\begin {equation}
\frac{g_{\rho \pi \pi}}{f_\rho}|_{VMD} = 1 \; .
\end {equation}
The disagreement between Eq.(11) and experiment:
\begin {equation}
\frac{g_{\rho \pi \pi}}{f_\rho}|_{EXP} = 1.21 \pm 0.02 \; ,
\end {equation}
provides strong support for the existence of radial excitations of the
rho-meson. Furthermore, the prediction from Eqs.(9)-(10) for the fitted
value $\beta \simeq 2.3$
\begin {equation}
\frac{g_{\rho\pi\pi}}{f_\rho}|_{QCD_{\infty}} = \frac{2}{\sqrt{\pi}}
\frac{\Gamma(\beta- 1/2)}{\Gamma(\beta-1)} \simeq 1.2 \; ,
\end {equation}
provides additional strong support for Dual-$QCD_\infty$.\\

Turning to the time-like region ($s>0$), clearly the zero-width feature
of the spectrum in $QCD_{\infty}$ is unrealistic. It is possible, though,
to make the pion form factor in Dual-$QCD_{\infty}$ compatible with
unitarity. In fact, unitarization of  Dual Model three-point functions
does not lead to unwanted consequences \cite{CAD3},\cite{LU}, as is the
case with n-point functions ($n \geq 4$) \cite{VEN}. It should be stressed
that very little change in the behaviour of $F_\pi(s)$ in the space-like
region is expected from smearing the spectrum in the time-like region.
This process involves, in addition to the parameter $\beta$ in Eq.(3),
at least one more free parameter related to the widths of the radial
excitations, as discussed next. The imaginary part of  $F_\pi(s)$ which
follows from Eq.(6) is
\begin{equation}
Im \; F_{\pi}(s) = \frac{\Gamma(\beta-1/2)}{\alpha' \sqrt{\pi}} \;
\sum_{n=0}^{\infty} \; \frac{(-1)^{n}}{\Gamma(n+1)} \;
 \frac{1}{\Gamma(\beta-1-n)} \; \pi \; \delta(M_n^2-s) \;.
\end{equation}
A simple unitarization prescription is to make the substitution
\cite{CAD3}, \cite{LU}
\begin{equation}
\pi \delta(M_{n}^{2} -s) \rightarrow \frac{\Gamma_{n} M_{n}}
{[(M_{n}^{2} - s)^{2} + \Gamma_{n}^{2} M_{n}^{2}]} \; .
\end{equation}
This Breit-Wigner resonance spectrum is expected to be a reasonable
approximation for relatively narrow resonances. In addition, the correct
threshold behaviour of the imaginary part of  $F_\pi(s)$ must be
enforced; in the chiral $SU(2) \times SU(2)$ limit
$Im \;{F_\pi(s)} \rightarrow s \rightarrow 0 $. As to the n-dependence of
the radial excitation widths, these are expected to grow with increasing
mass. The folowing simple-minded ansatz will be adopted for the
purpose of illustration: $\Gamma_n = \gamma \; M_n$, with
$\gamma$ fixed by the rho-meson parameters ($\gamma \simeq 0.2$). Having
thus specified the imaginary part of  $F_\pi(s)$, its real part can be
computed from the dispersion relation
\begin{equation}
Re \; F_{\pi}(s) = \frac{1}{\pi} \;\int_{- \infty}^{+ \infty} \;
\frac{Im \; F_{\pi}(s')}{(s' - s)} \;ds' \; ,
\end{equation}
where the integral is to be understood as its principal value.\\

It has been checked that the nice agreement between $F_\pi(s)$ in
Dual-$QCD_\infty$ and the data for $s \leq  0$ remains
essentially unchanged by unitarization. However, once smeared,
$F_\pi(s)$ may be used to compute the vector spectral function defined
as the imaginary part of the two-point function
\begin{eqnarray*}
\Pi_{\mu \nu}^{VV} (q^2) = i \; \int \; d^4 \; x \; e^{i q x} \; \;
 <0|T(V_{\mu}(x) \; \; V_{\nu}^{\dagger}(0))|0> \; 
\end{eqnarray*}
\begin{equation}
 = \; (- g_{\mu \nu} \; q^{2} + q_{\mu} q_{\nu}) \; \Pi_{V} (q^{2}) \; ,
\end{equation}
where $V_{\mu}(x) = :\bar{q}(x)  \gamma_{\mu}  q(x):$, and $q=(u,d)$. In
fact, using the normalization in which $\lim_{ s \rightarrow \infty}
\rho_V(s) =  1/2$ (at the one loop level in QCD), with $\rho_{V}(s) 
\equiv \frac{1}{\pi} \;Im \Pi_V(s)$,
one has in the chiral limit
\begin{equation}
|F_\pi(s)|^2 = 12 \; \rho_V(s) \; .
\end{equation}
Figure 2 shows the ALEPH data on the vector spectral function together
with that computed from the unitarized pion form factor (solid curve).
It is quite encouraging that despite the simplicity of the unitarization
ansatz chosen here, the agreement with
the data is quite reasonable up to about $ s \simeq 1.3 \;\mbox{GeV} ^2$.
Globally, the agreement is even better, viz. the area under the
theoretical spectral function is
\begin{equation}
\int_{0}^{s_0}\; \rho_V(s)|_{QCD_\infty} \; ds \;= 1.1 \; \mbox{GeV}^2 \; ,
\end{equation}
to be compared with the experimental value
\begin{equation}
\int_{0}^{s_0}\; \rho_V(s)|_{EXP} \; ds \;= 0.9 \; \mbox{GeV}^2 \; ,
\end{equation}
where $s_0 = 1.25 \; \mbox{GeV}^2$. For higher energies the Dual-$QCD_\infty$
spectral function falls faster than the data. However, this is not a
serious drawback, since at these energies one expects perturbative QCD
to take over. In fact, standard models of the spectrum consist of
hadronic resonance parametrizations up to $s_0$, and from there onwards
the perturbative QCD expression.\\

In summary, Dual-$QCD_\infty$ provides a reasonable model for the
spectrum of infinite zero-width vector-isovector resonances, to the extent
that the resulting pion form factor is in very good agreement with 
experimental data in the space-like region, after fitting
the single free parameter of the model ($\beta$ in Eqs. (3) and (6)).
Moreover, the resulting pion mean squared radius, and the ratio of strong
to electromagnetic couplings also  agree very well with the data.
In contrast, naive VMD is in poor agreement with experiment on all three
counts. In the time-like region, the model can be easily unitarized.
A very simple ansatz shows that good agreement with the experimental vector
spectral function can be achieved, both locally and globally, up to about
$s \simeq 1.3 \;\mbox{GeV}^2$. This is normally sufficient, as for higher
energies one usually assumes the spectral function to be given by
perturbative QCD. Clearly, the unitarization ansatz discussed here is
subject to considerable improvement.

\begin{center}
{\bf Figure Captions}
\end{center}

Figure 1. Dual-$QCD_\infty$ pion form factor, Eq.(6), in the space like
region for the fitted parameter $\beta \simeq 2.3$ (solid curve), together
with naive VMD ($\beta= 2$) (dash curve), and the experimental data
\cite{DATA}-\cite{AMEN}.

Figure 2. Dual-$QCD_\infty$ vector spectral function in
the time-like region, from Eq.(18) after unitarization, together with the
ALEPH data \cite{ALEPH}.

\newpage
\begin{figure}[tp]
\epsffile{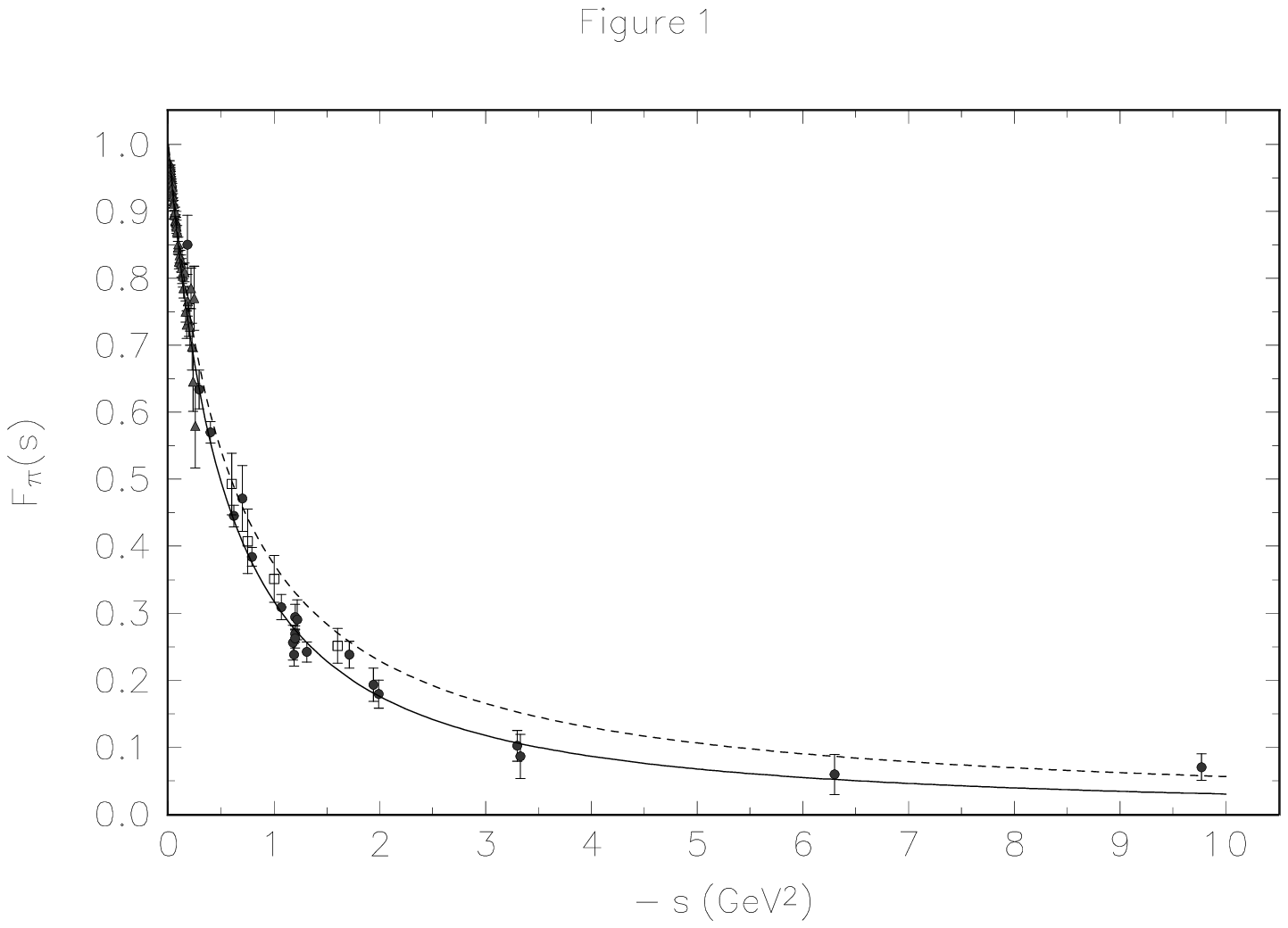}
\caption{}
\end{figure}
\newpage
\begin{figure}[tp]
\epsffile{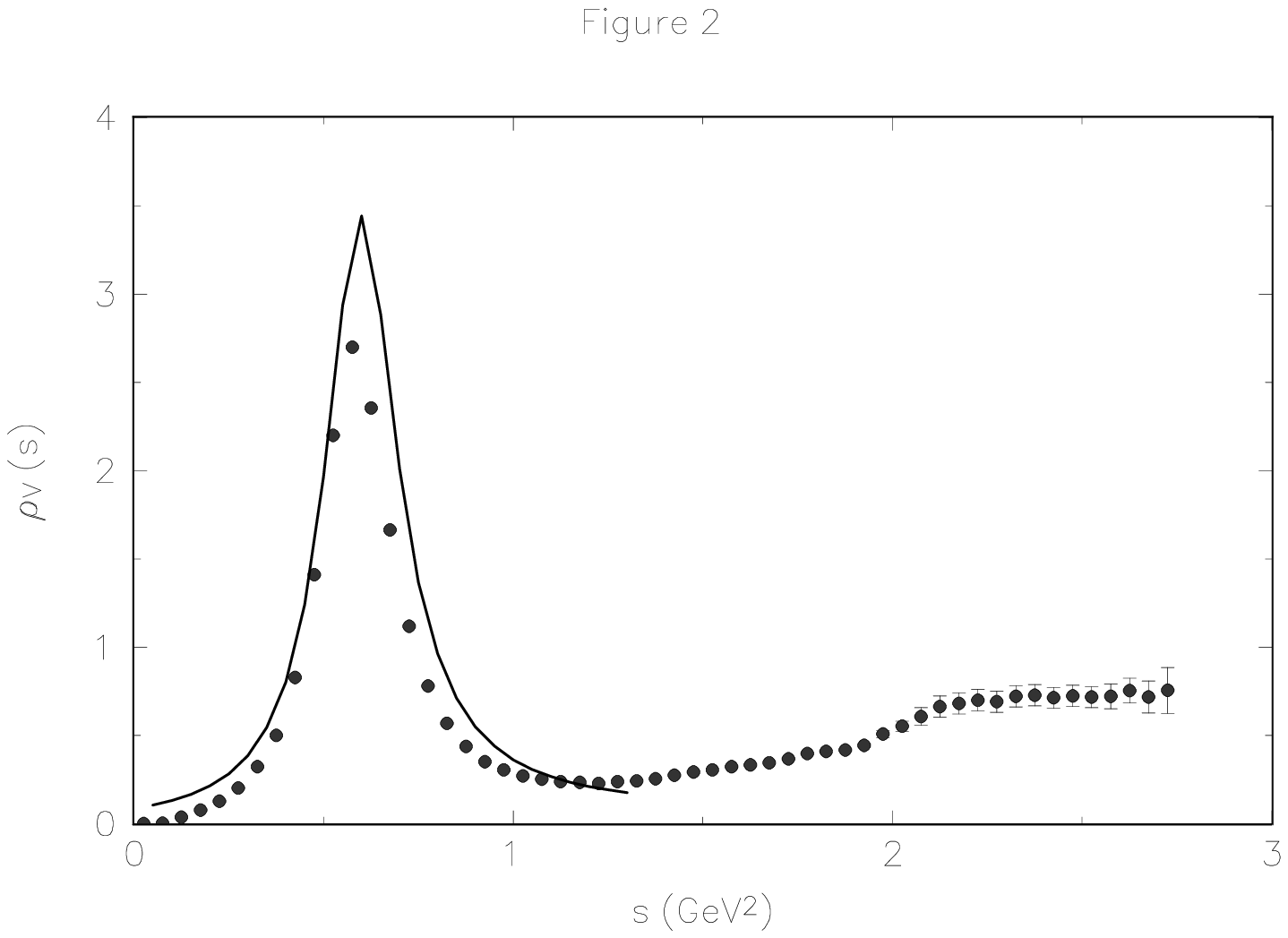}
\caption{}
\end{figure}

\end{document}